\documentclass[aps,pra,floatfix,twocolumn,showpacs,tightenlines,groupedaddress,superscriptaddress,amsmath]{revtex4}
\usepackage{verbatim}

\pdfoutput=1

\ifx\pdftexversion\undefined
  \usepackage[dvips]{color,graphicx}
\else
  \usepackage[pdftex]{color,graphicx}
\fi

 \usepackage{ifpdf}
    \ifpdf
 	 \usepackage[pdftex]{color,graphicx}
    \else
  \usepackage[dvips]{color,graphicx}
 \fi

\def	\bse{\begin{subequations}}
\def	\ese{\end{subequations}}

\def \be{\begin{equation}}
\def \ee{\end{equation}}
\def \bew{\begin{widetext}\begin{equation}}
\def \eew{\end{equation}\end{widetext}}
\def \bmlett{\begin{mathletters}}
\def \emlett{\end{mathletters}}

\def \omegam{\omega_M}

\def \nbar{\bar{n}_{\rm dr} }
\def \nth{\bar{n}_{\rm th}}

\def \tm{\tilde{m}}

\def	\neff{\bar{n}_{\rm eff}}

\def \ua{\uparrow}
\def \da{\downarrow}
\def \ra{\rightarrow}

\def \hrho{\hat{\rho}}
\def \hx{\hat{x}}

\def \hA{\hat{A}}
\def \hF{\hat{F}}
\def \hI{\hat{I}}

\def \hX{\hat{X}}
\def \hY{\hat{Y}}

\def \ha{\hat{a}}
\def \hb{\hat{b}}
\def \hc{\hat{c}}
\def \hH{\hat{H}}

\def \hn{\hat{n}}
\def \hm{\hat{m}}

\def \Pdr{P}

\begin{document}

\title{Full counting statistics of energy fluctuations in a driven quantum resonator}
\author{A. A. Clerk}
\affiliation{Department of Physics, McGill University, Montreal,
Quebec, Canada H3A 2T8}

\date{May 31, 2011}

\begin{abstract}
We consider the statistics of time-integrated energy fluctuations of a driven bosonic resonator (as measured by a QND detector), using the standard Keldysh
prescription to define higher moments.  We find that due to an effective cascading of fluctuations, 
these statistics are surprisingly non-classical:  the low-temperature, quantum
probability distribution is not equivalent to the high-temperature classical distribution evaluated at some effective temperature.  Moreover, 
for a sufficiently large drive detuning and low temperatures, the Keldysh-ordered quasi-probability distribution characterizing these fluctuations 
fails to be positive-definite; this is similar to the full counting statistics of charge in superconducting systems.  
We argue that this indicates a kind of non-classical behaviour akin to that tested by Leggett-Garg inequalities.
\end{abstract}

\maketitle



\section{Introduction}

The statistics of photon fluctuations in various setting is by now an almost textbook phenomena.  Most familiar are the statistics that would be measured by a photodetector.  Relatively less attention has been paid to photon fluctuations in the case where the detection is done in a non-demolition manner, 
meaning that energy quanta are measured without destroying them. The study of such fluctuations is not just a theoretical curiousity, as 
quantum non-demolition (QND) photon detection is experimentally feasible both optically \cite{Grangier98}, as well as in both cavity QED systems \cite{Haroche99, Guerlin07, Gleyzes07} and superconducting circuit-QED systems \cite{Schuster07, Johnson10}. 
In the latter systems, one uses dispersive interactions to detect photon number inside a cavity.  
QND detection of phonon number in a mechanical resonator may also soon be possible in optomechanical systems \cite{Harris08, Sankey10}, where the energy of a mechanical resonator is directly coupled to the frequency of an optical cavity.

Motivated by developments in optomechanics, we recently investigated the low-frequency energy fluctuations of a driven, damped harmonic resonator, focusing on the possibility of measuring these fluctuations non-destructively using an optomechanical cavity \cite{Clerk10a}.   In the zero-temperature, quantum limit, the instantaneous state of such a resonator is simply a coherent state, yielding a Poissonian distribution of phonon number.  Our focus was instead on understanding how the mechanical phonon number $\hat{n}$ fluctuated in time.  
These fluctuations are characterized by a power spectral density $S_{nn}[\omega]$, or equivalently by the second central moment of the time-integrated phonon number $\hat{m}$:
\begin{eqnarray}
	\hat{m} & \equiv & \int_0^t dt' \hat{n}(t').  
	\label{eq:mDefn}
\end{eqnarray}

As could be anticipated, both $S_{nn}[\omega]$ and $\langle (\delta \hat{m})^2 \rangle = \langle \hat{m}^2 \rangle - \langle \hat{m} \rangle^2$ have a low-frequency ``shot-noise" term proportional to the average number of phonons induced by the drive, $\nbar$; detecting this shot noise contribution would be direct evidence for the quantization of the mechanical resonator's energy.  As the quantum signature here scales as $\nbar \gg 1$, measuring these low-frequency energy fluctuations is an easier way of detecting quantum behaviour than attempting to resolve the instantaneous phonon number and
individual quantum jumps.  Our study also addressed the non-Gaussian nature of the driven energy fluctuations by calculating the third central moment 
$\langle (\delta \hat{m})^3 \rangle$; surprisingly, we found that while this quantity is always positive classically, it could become negative in the low-temperature quantum limit.  As such, the third moment is far more sensitive to classical-quantum differences than the second moment.  

To fully understand the significance of this result, one needs to consider the full probability distribution characterizing the low-frequency fluctuations of $\hn$, 
and compare its form in the classical and quantum limits.  This is the objective of this paper:  we calculate the distribution $P(m)$ in the long-time limit using the standard Keldysh operator ordering \cite{Levitov93, Levitov96, Nazarov03b}.
We find that the anomalous negative value of $\langle (\delta \hat{m})^3 \rangle$ results from a kind of cascaded fluctuation effect 
\cite{Nagaev02, Beenakker03}, which can be heuristically attributed to a correlated fluctuation in the resonator temperature.  We also find that this negative skewness is a precusor of something rather dramatic:  {\it in the quantum limit, the fluctuations of $m$ are most naturally described by a quasi-probability distribution $P(m)$ which is not positive definite}.  Such negative counting statistics have been encountered before in the study of charge transfer in superconducting systems \cite{Belzig01, Nazarov03b}; their interpretation requires some care.  As we discuss in some detail, they are indicative of non-classical temporal correlations, and are thus somewhat similar to having violated a Leggett-Garg inequality \cite{Leggett85}.  Detecting these effects thus represents a new way of detecting non-classical behaviour in a driven quantum resonator.

The remainder of this paper is structured as follows.  In Sec.~\ref{sec:Model}, we introduce our basic model, 
and present our main results for the generating function of $P(m)$; we also give a compact review of the Keldysh ordering of higher moments for those not familiar with this topic.  In Sec.~\ref{sec:Classical},  we discuss the form of the distribution in the classical limit.  Sec.~\ref{sec:Quantum} is dedicated to the distribution in the quantum limit, while Sec.~\ref{sec:Interpretation} is devoted to interpreting the negative quasi-probabilities which emerge.  Finally, in Sec.~\ref{sec:Measurement} we discuss issues related to the measurement of these effects.  An appendix is included which shows how the Keldysh operator ordering emerges naturally in the proposed experimental realization of Ref.~\cite{Clerk10a}, where $m$ is measured by using homodyne interferometry to detect the frequency shift of an auxiliary cavity.
Finally, we note that a fermionic analogue of the present problem, the full counting statistics of electronic charge fluctuations in a chaotic quantum dot, were studied in Ref.~\cite{Pilgram03}.

\section{Model and calculation}
\label{sec:Model}

\subsection{Statement of the problem}
\label{subsec:Intro}

Our damped, driven harmonic resonator is described by the Hamiltonian:
\begin{eqnarray}
	\hH \equiv \hH_0 + \hH_{\gamma} =  \hbar \omegam \hc^\dag \hc   -  \hbar f \left( e^{i \omega_D  t} \hc + h.c. \right)  + \hH_{\gamma} .
	\label{eq:Hamiltonian}
\end{eqnarray}
Here, the first term describes the resonator (frequency $\omegam$, number operator $\hn = \hc^\dag \hc$), $\hH_{\gamma}$ describes the damping (at a rate $\gamma$) and heating of the oscillator by a thermal bath, and $f$ is the magnitude of the coherent oscillator driving force (frequency $\omega_D = \omegam + \delta$).  We take $H_{\gamma}$ to correspond to  the standard model of a linear coupling to an Ohmic oscillator bath.  We also define the dimensionless oscillator force susceptibility as:
\begin{eqnarray}
	\chi & = &	\frac{1}{1 + 4 (\delta / \gamma)^2}.
	\label{eq:ChiDefn}
\end{eqnarray}

We will be interested in the statistics of the time-integrated energy $\hm$ (c.f.~Eq.~(\ref{eq:mDefn})) in the case where the oscillator has equilibrated to both the driving force and thermal bath long before the initial time $t=0$.  We will also focus exclusively on the long time limit, e.g.~an integration time $t$ which is long compared to $1 / \gamma$.  The average and second moment of $\hm$ are easily found by solving the Heisenberg-Langevin equations for our system \cite{Gardiner00,ClerkRMP}.  In the long time limit, the average $\langle \hm \rangle  \sim  \left[ \nbar + \nth \right] t$,
where $\nth$ denote the thermal number of oscillator quanta (determined by the bath temperature), and $\nbar = (2 f/ \gamma)^2 \chi$ is the average number of quanta due to the driving force.

For the second central moment, we find in the long time limit:
\begin{eqnarray}
	\langle (\delta \hm)^2 \rangle & \sim & \langle (\delta \hm)^2 \rangle_{\rm dr} +  \langle (\delta \hm)^2 \rangle_{\rm th},
	\label{eq:SecondMoment}
\end{eqnarray}
where
\begin{eqnarray}
	\langle (\delta \hm)^2 \rangle_{\rm th} & = & 
		\frac{ 2 \nth (1+ \nth) t}{\gamma}  
\end{eqnarray}
represents a purely thermal contribution whereas
\begin{eqnarray}
	\langle (\delta \hm)^2 \rangle_{\rm dr} & = & 
		 \frac{ 8 \nbar \chi  t} {\gamma}   \left(\nth + \frac{1}{2} \right)
	\label{eq:DriveVariance} 
\end{eqnarray}
represents extra energy fluctuations due to the driving force.  The last term in Eq.(\ref{eq:DriveVariance}) here survives in the limit of zero temperature
(i.e.~$\nth \ra 0$), and is a quantum effect: it corresponds to the shot noise fluctuations arising from the discreteness of the resonator's energy.  

\subsection{Higher moments and the Keldysh ordering}

Before calculating higher moments and the full distribution of $\hm$, we must pause to consider the operator-ordering ambiguity arising from the non-commutativity of $\hn(t)$ at different times.  In calculating the second moment, we have naively defined the variance as $\langle \hm^2 \rangle$, an expression which is naturally symmetrized in terms of $\hn$ products, i.e.:
\begin{eqnarray}
	\langle \hm^2 \rangle & = & 
		\int_0^t dt_1 \int_0^t dt_2 \langle \hn(t_1) \hn(t_2) \rangle  \nonumber \\
		& = & 
		\frac{1}{2} \int_0^t dt_1 \int_0^t dt_2 \langle \{ \hn(t_1) ,  \hn(t_2) \} \rangle.
		\label{eq:SecondMoment}
\end{eqnarray}
If we define higher moments in the same way (e.g. define the $j$th moment to be $\langle \hm^j \rangle$), they too would be naturally symmetrized; one might expect that this is then a sensible way to proceed.  

Unfortunately, being sensible is not enough to guarantee physical relevance: similar to the standard theory of photodetection \cite{Glauber63}, one must instead model the actual detection scheme to properly understand how the measured moments correspond to a given operator ordering.  As we are interested in non-destructive detection, the answer here {\it will not} be the normal-ordering prescription used in photodetection.  A similar problem arises in the measurement of current fluctuations in quantum coherent conductors;  the answer emerging from studies of this question is the so-called Keldysh operator ordering.  This ordering appears naturally in a number of idealized measurement setups \cite{Levitov93, Levitov96, Nazarov03b}; it can also be given an elegant motivation using a path-integral formulation of the Keldysh field-theoretic technique \cite{Kamenev05, Kamenev09}.  For the second moment, the ordering coincides with the simple definition in Eq.~(\ref{eq:SecondMoment}); for higher moments, the ordering prescriptions have no simple intuitive form (see, e.g., Ref.\cite{Clerk10a} for the explicit form for the third moment).
%

 \subsection{Using an auxiliary qubit to obtain $P(m)$}
 \label{subsec:QubitModel}

We give here a quick derivation of the Keldysh ordering, and use it to derive $P(m)$ for our system.
Following Ref.~\cite{Levitov96}, we consider an idealized method for measuring $P(m)$, in which $\hn$ couples dispersively to the $\hat{\sigma}_z$ operator of an auxiliary two-level system (TLS) with a coupling strength $k/2$:
\begin{eqnarray}
	\hH_{\rm int} = \frac{\hbar k}{2}  \hn \hat{\sigma}_z .
	\label{eq:HInt}
\end{eqnarray}
As there are no other terms involving the TLS in the Hamiltonian, we see that it simply experiences a magnetic field $\propto \hn$.  If $\hn$ were just a classical, time-dependent field $n(t)$, then during the time interval between $0$ and $t$, the TLS would precess an angle $\theta = k \int_0^t n(t') dt' = k m$.  If now $m$ was a classically stochastic variable described by the distribution $P(m)$, then the average of $e^{-i \theta}$ over this distribution (at a fixed coupling $k$) would be:
\begin{eqnarray}
	\langle e^{-i \theta} \rangle_k = \int dm P(m) e^{-i k m} \equiv \Lambda[k].
	\label{eq:PhaseAverage}
\end{eqnarray}
Thus, when viewed as a function of $k$, the average of the precession phase directly yields the moment generating function $\Lambda[k]$ of the distribution $P(m)$. 

The above correspondence now provides a means for defining $P(m)$ in the quantum case \cite{Levitov96}: we simply use the fact that the average on the LHS of Eq.~(\ref{eq:PhaseAverage}) corresponds to  $\rho_{\ua \da}(t) / \rho_{\ua \da}(0)$, where $\rho_{\ua \da}(t)$ is an off-diagonal matrix element of the TLS's reduced density matrix.  We can thus {\it define} the moment generating function $\Lambda[k]$ (and hence $P(m)$) in the quantum case via:
\begin{eqnarray}
	\Lambda[k] \equiv 
		\frac{ \rho_{\ua \da}(t) }{  \rho_{\ua \da}(0)} \Bigg|_k 
		=
		\textrm{Tr }_{\rm sys} \left[
			\hat{U}(t;k) \hrho_{\rm sys} \left( \hat{U}(t;-k) \right)^{\dag} \right]	,
	\label{eq:TLSKeldyshDefn}
\end{eqnarray}
where the time evolution operator $\hat{U}$ is defined as:
\begin{eqnarray}
	\hat{U}(t;k) & = &
		\mathcal{T} \exp\left[ -\frac{i}{\hbar} \int_0^t \left( \hH(t') + \frac{\hbar k}{2} \hn(t') \right)  \right].
\end{eqnarray}
Here, $\hH$ is given in Eq.~(\ref{eq:Hamiltonian}), $\hrho_{\rm sys}$ is the initial measured system (i.e.~cavity plus bath) density matrix, and the trace is taken over all  system degrees of freedom.  Further, the symbol $\mathcal{T}$ denotes time ordering.
Eq.~(\ref{eq:TLSKeldyshDefn}) uniquely specifies the operating ordering to use for each moment of $P(m)$; this is the Keldysh ordering.  We stress that the same ordering emerges in the analysis of other idealized measurement setups \cite{Nazarov03b}; we also show in Appendix A that it applies to a realistic setup where $\hn$ is coupled dispersively to a detector cavity whose frequency is monitored using homodyne detection.  

  
In our case, the above scheme not only motivates the Keldysh ordering, it also gives us a convenient way to calculate the generating function $\Lambda[k]$.  The reduced density matrix $\hrho$ describing the TLS and the driven resonator (i.e.~only the resonator's dissipative environment is traced out) obeys the following standard master equation:
\begin{eqnarray} 
	\dot{\hrho} &=&
		-\frac{i}{\hbar} \left [ \hH_0, \hrho \right ]
	+ \gamma (\nth+1) \mathcal{D}[\hc] \hrho
	+ \gamma \nth
	\mathcal{D}[\hc^\dagger] \rho,
	\label{eq:Master}
\end{eqnarray}
where $H_0$ is defined in Eq.~(\ref{eq:Hamiltonian}), and where for any operator $\hA$ we define 
$	\mathcal{D}[\hA] \hrho = 
		\hA \hrho \hA^{\dag} -  \left(
			\hA^{\dag} \hA \hrho + \hrho \hA^{\dag} \hA
		\right)/2 $.
We are using the ``quantum optics"
version of the master equation, which is appropriate for the high-$Q$ limit we consider.

As shown in Ref.~\cite{Clerk07}, using standard phase space techniques, one can solve Eq.~(\ref{eq:Master}) and thus directly obtain $\Lambda[k]$.  
Ref.~\cite{Clerk07} used this quantity to study dephasing and coherence revivals of the TLS; the emphasis was on understanding $\rho_{\ua \da}(t)$ for a fixed value of the coupling $k$.  In contrast, our focus here is on how $\rho_{\ua \da}(t)$ behaves as a function of $k$ in the long-time limit, as it is this behaviour which will determine $P(m)$ in the long-time limit. 

\subsection{Generating function for P(m)}

Using the procedure described above, and taking the long-time limit, the final result for the moment generating function $\Lambda[k]$ has the simple form 
(c.f.~Eq.~(22) in Ref.~\cite{Clerk07}):
\begin{eqnarray}
	\Lambda[k] = \Lambda_{\rm dr}[k] \Lambda_{\rm th}[k]
\end{eqnarray}
where $\Lambda_{\rm th}[k]$ describes a purely thermal (drive-independent) contribution, and $\Lambda_{\rm dr}[k]$ describes additional fluctuations related to the drive.  One finds:  
\begin{subequations}
\begin{eqnarray}
	\Lambda_{\rm dr}[k] & = & 
		\exp \left[ \frac{-i k \nbar t}{1 + 4 i  \chi \left(\nth + \frac{1}{2} \right) (k/\gamma) - \chi (k/ \gamma)^2 } \right],
		\label{eq:MGF} 
		\\
	\Lambda_{\rm th}[k] & = & 
	\exp{ \left[ -\frac{\gamma t}{2}
			\left(
				\sqrt{ 1 + 4 i \left( \nth + \frac{1}{2} \right)\frac{k}{\gamma} -\frac{k^2}{\gamma^2}  } - 1
			\right) \right] }. \nonumber \\
		\label{eq:MGFTherm} 
\end{eqnarray}
\end{subequations}
In taking the long-time limit, we have simply dropped terms in $\Lambda[k]$ which decay exponentially in time as $\exp(-\gamma t / 2)$ or faster.

Recalling that the $j$th cumulant of $m$ is given by $i^j \frac{d^j}{d k^j} \ln \Lambda[k] \Big |_{k=0}$, 
one can easily check that Eqs.~(\ref{eq:MGF}) and (\ref{eq:MGFTherm}) yield the same second and third cumulants obtained from the Heisenberg-Langevin approach.  We see that the purely thermal fluctuations described by $\Lambda_{\rm th}$ are independent of the additional drive-induced fluctuations described by $\Lambda_{\rm dr}$; further note that these purely thermal fluctuations vanish in the limit of zero temperature.  Thus, in the remainder of the paper we focus on the case $\nbar \gg \nth,1$, and thus focus attention to the distribution $\Pdr(m)$ generated by $\Lambda_{\rm dr}[k]$.

%
%
%

\section{Energy fluctuation statistics in the classical limit}
\label{sec:Classical}

To gain some intuition, it is useful to first consider the driven energy counting statistics in the classical, high-temperature limit.   Formally, one transforms the distribution described by Eq.~(\ref{eq:MGF}) to a distribution describing the time-integrated energy $s = \int_0^t dt' E(t') = \hbar \omegam m$,  One can then rigorously take the $\hbar \ra 0$ limit.  Transforming back to our original variable $m$, one finds:
\begin{eqnarray}
	\Lambda_{\rm dr, cl}[k] = 
		\exp \left[ \frac{-i k \nbar t}{1 + 4 i  \chi \nth (k/\gamma)  } \right].
		\label{eq:MGFclass}
\end{eqnarray}
In the long time-limit of interest, the corresponding probability distribution function can be found within a saddle point-approximation, yielding:
\begin{eqnarray}
	P_{\rm cl}(m) \simeq 
		\frac{1}{\sqrt{8 \pi  m \sigma_{\rm cl}^2 }  }  \cdot 
		\exp \left[
			\frac{ -\left(  \sqrt{m} - \sqrt{\nbar t} \right)^2}{2 \sigma_{\rm cl}^2}  
		\right],
		\label{eq:Pclass}
\end{eqnarray} 
with 
\begin{eqnarray}
	 \sigma_{\rm cl}^2 = 2 \nth \chi /  \gamma.  
	 \label{eq:ClassVariance}
\end{eqnarray}
We see that classically, $m$ is well-approximated as being the square of Gaussian random variable with mean $\sqrt{\nbar t}$ and standard deviation $\sigma_{\rm cl}$.  This of course implies that even in the classical limit, $m$ is not itself a Gaussian variable.

The above behaviour is easily understood.  Writing the complex cavity amplitude $a(t)$ in terms of its mean value
$\sqrt{\nbar} e^{-i \omega_d t}$ and a thermally fluctuating part
$ \delta a(t)$, we have:
\begin{eqnarray}
	m(t) & = & \int_0^t dt' \left| \sqrt{\nbar} + e^{i \omega_d t'} \delta a(t') \right|^2 
		 \nonumber \\
	& = & 
		\int_0^t dt' \left| \sqrt{\nbar} +\frac{ \delta X(t) + i \delta Y(t) }{\sqrt{2}} \right|^2. \label{eq:Classm}
\end{eqnarray}
In the second line, we have written the fluctuation $\delta a(t)$ in terms of 
real-valued quadratures $\delta X(t)$ and $\delta Y(t)$.  In the large $\nbar$ limit, only the low-frequency part of the ``intensity-quadrature" noise $\hX$ will be enhanced by $\nbar$.  To a good approximation, we may thus drop the $\hY$ contribution, and replace $\hX$ by its low frequency part.  We thus have:  
\begin{eqnarray}
	m(t) & \simeq &
		\left( \sqrt{\nbar t}  +  \frac{1}{\sqrt{2 t}} \int_0^t dt' \delta X(t')  \right)^2 
		\nonumber \\
		& \equiv & 
			 \left( \sqrt{\nbar t}  +  \overline{\delta X}(t) \right)^2.
			 \label{eq:Classm2}
\end{eqnarray}
One can easily confirm that the time-averaged intensity-quadrature noise $\overline{\delta X}(t)$ defined above is a Gaussian random variable with zero mean and variance $2 \nth \chi / \gamma $; Eq.~(\ref{eq:Classm2}) is thus in agreement with Eq.~(\ref{eq:Pclass}).  
It is worth noting that a full classical calculation of 
$P(m)$ (including purely thermal effects) yields an answer in complete agreement with the classical limit of the moment generating functions given in Eqs.~(\ref{eq:MGF}) and (\ref{eq:MGFTherm}).

\section{Energy fluctuation statistics in the quantum regime}
\label{sec:Quantum}

\subsection{Basic results}

It is tempting to make a simple extrapolation of the classical energy statistics to the quantum regime.  Again, in the large 
$\nbar$ limit it is the amplification of the thermal intensity-quadrature fluctuations $\overline{\delta X}(t)$ (cf.~Eq.~(\ref{eq:Classm2})) which determine $P(m)$;
one might expect that the only difference in the quantum case is that these quadrature fluctuations are driven by both thermal and zero-point force noise.   We would thus expect the distribution to again be given by Eq.~(\ref{eq:Pclass}), with the simple modification that the variance $\sigma_{\rm cl}$ in Eq.(\ref{eq:ClassVariance}) should be increased to include zero-point fluctuations via the substitution $\nth \ra \nth + 1/2$.  

However, as already mentioned in the introduction, this is not the case.  Instead, the full quantum moment generating function in Eq.~(\ref{eq:MGF}) is related to the classical one (c.f.~Eq.~(\ref{eq:MGFclass})) by the simple substitution:
\bse
\begin{eqnarray}
	\Lambda_{\rm dr}[k; \nth] & = &  \Lambda_{\rm dr, cl}\left[k; \nth \rightarrow \bar{n}_{\rm eff} [k]  \right],  \\
	\bar{n}_{\rm eff}[k] & = & \nth + \frac{1}{2} + i \frac{k}{4 \gamma}.
	\label{eq:neff}
 \end{eqnarray}
\ese
Thus, one shifts $\nth$ both by the constant $1/2$ (reflecting the inclusion of zero point force noise), as well as by an imaginary, $k$-dependent term.

The $k$-dependence of $\neff$ implies non-trivial quantum corrections to the third cumulant and higher involving a kind of feedback, whereby higher-order cumulants depend on the form of lower-order cumulants.  In the case of the third cumulant $\langle \langle m^3 \rangle \rangle$, one finds:
 \begin{eqnarray}
	\langle \langle m^3 \rangle \rangle & = &
			\langle \langle m^3 \rangle \rangle_{\rm cl'} 
				- 3 \left[ \frac{d}{d \nth} \langle \langle m^{2} \rangle \rangle_{\rm cl'} \right] \frac{1}{4 \gamma},
	\label{eq:MomentFeedback}
\end{eqnarray}
where we use $\langle \langle m^j \rangle \rangle_{\rm cl'}$ to denote the naive expectation for the $j$th cumulant:  the $j$th classical 
cumulant obtained from Eq.~(\ref{eq:MGFclass}), with the substitution $\nth \ra \nth + 1/2$.

Heuristically, this feedback of lower moments into higher moments is analogous to the situation in so-called  ``cascaded" Langevin approaches \cite{Nagaev02, Beenakker03}.  One could heuristically obtain the feedback term in Eq.~(\ref{eq:MomentFeedback}) using such an approach, starting with the assumption that the effective thermal occupation $\nth$ in the classical distribution fluctuates in a way that is driven by (and hence correlated with) $\delta m$.
Assuming that the fluctuations of $\delta \nth$ are slow compared to those of $\hn$ (allowing a two-step averaging procedure), one 
obtains:
\begin{eqnarray}
	\langle \langle m^3 \rangle \rangle & = &
			\langle \langle m^3 \rangle \rangle_{\rm cl'} 
				+ 3 
				\left[ 
					\frac{d  }{d \nth}  \langle \langle m^{2} \rangle \rangle_{\rm cl'}
				\right] 
				\langle \delta \nth(t) \cdot \delta m(t) \rangle.
				\nonumber \\
\end{eqnarray}
This recovers Eq.~(\ref{eq:MomentFeedback}) if we take $\langle \delta \nth(t) \cdot \delta m(t) \rangle = - 1/ 4 \gamma$
While usually derived in a heuristic fashion, cascaded Langevin approaches have been used successfully to understand higher cumulants of current fluctuations in electronic conductors \cite{Nagaev02, Beenakker03}.  Here, we stress that this picture emerges directly from a fully quantum calculation.

Turning to the explicit form of the third cumulant, evaluating Eq.~(\ref{eq:MomentFeedback}) yields:
 \begin{eqnarray}
	\langle \langle m^3 \rangle \rangle & = &
		\frac{\nbar t}{\gamma^2} \chi^2 \left(
			24 \left( 1 + 2 \nth \right)^2  
				- \frac{6 }{\chi} 
			\right).
	\label{eq:QuantumThirdMoment}
\end{eqnarray}
The first term here is just the expected classical answer; it is always positive.
The second term here is the non-trivial ``feedback" quantum correction; as it involves the second moment of the classical distribution, it is lower-order in the susceptibility $\chi$ than the first term.  As a result, this correction can make the skewness negative for a sufficiently small $\chi$, something that is impossible classically.  Further, in the limit of a strongly-detuned drive (i.e.~$\chi \ll 1$), the quantum skewness has a much larger magnitude (by a factor $1/\chi$) than the corresponding classical answer.  The quantum ``feedback" corrections similarly enhance all higher moments over the corresponding classical answer in the large detuning limit.

\subsection{Negative probabilities at large drive detuning}

\begin{figure}[t]
	\centering
	\includegraphics[width= 0.91 \columnwidth]{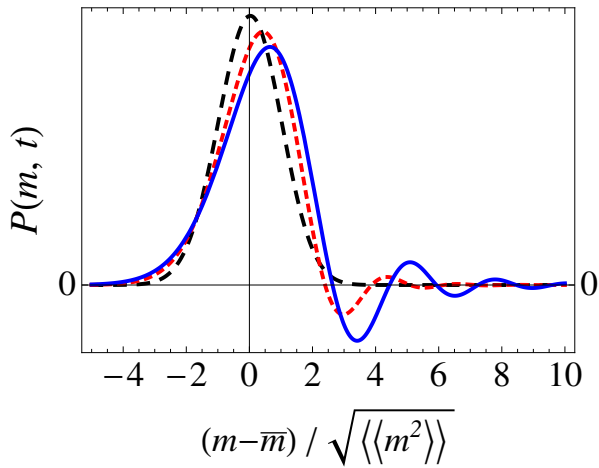}
	\caption{
		The distribution $P(m)$ of integrated energy fluctuations $m$ of a driven resonator, evaluated at a time $t = 10 / \gamma$
		a drive detuning of $\delta = 10 \gamma$, 
		and a driving force strength which yields an average number of cavity quanta $\nbar = 5$.  The various curves correspond to different
		resonator temperatures:  $\nth = 3$ (black wide-dashed), $\nth = 0.25$ (red, small-dashed) and $\nth = 0$ (solid blue).  As discussed in
		the text, for large drive detunings and low temperatures, the distribution $P(m)$ fails to be positive definite.}	
	\label{fig:PmTemperature}
\end{figure}

The enhanced role of the non-trivial quantum corrections (arising from the $k$ dependence of $\neff[k]$) in the large detuning limit $| \delta | \gg \gamma$ 
is even more apparent if one looks at the form of the full distribution $\Pdr(m)$.  One finds that at sufficiently low temperature and large detuning, 
these quantum corrections lead to $\Pdr(m)$ becoming non-positive definite (see Fig.~\ref{fig:PmTemperature}).  This can be demonstrated analytically by just using the first four cumulants of $P(m)$.  The solution of the Hamburger moment problem \cite{HamburgerRef} is a necessary and sufficient set of conditions for a set of moments to correspond to a positive-definite probability distribution.   Letting  $C_j$ to denote the $j$th cumulant scaled by the variance (e.g. $C_3 = \langle\langle m^3\rangle \rangle / [\langle \langle m^2 \rangle \rangle]^{3/2}$),  the lowest-order Hamburger positivity constraint involving the third moment is:
\begin{eqnarray}
	\left( C_3 \right)^2  	& \leq &	 C_4 + 2. 
	\label{eq:Hamburger}
\end{eqnarray}
In the long time limit and for large detunings, $C_3 \propto \chi^{-1/2}$ while $C_4$ is independent of $\chi$.  The above constraint is thus violated by $\Pdr(m)$ for a sufficiently large detuning; in the large-$t$ limit, the condition for violation becomes:
\begin{eqnarray}
	(\delta/\gamma)^2  > \frac{8}{9} (\nbar t \gamma) \left(1 + 2\nth\right)^3. 
\end{eqnarray}
Heuristically, the quantum ``feedback" contribution to the third moment (second term in Eq.~(\ref{eq:MomentFeedback})) 
has made it too large for $\Pdr(m)$ to be positive definite  
\footnote{We stress that Eq.~(\ref{eq:Hamburger}) is a sufficient but not necessary condition for $\Pdr(m)$ to exhibit negativity; higher-order Hamburger constraints are violated at even smaller values of $|\delta|$.}.  

\begin{figure}[t!]
	\centering
	\includegraphics[width= 0.91 \columnwidth]{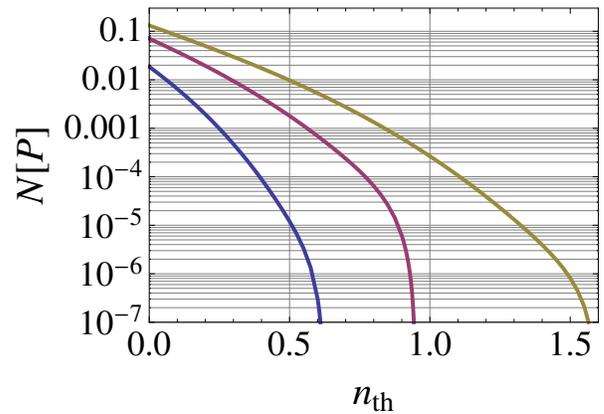}
	\caption{
		The integrated total negativity of the distribution $P(m)$ as a function of temperature $\nth$, for
		a time $t = 10 / \gamma$ and a driving force amplitude which yields $\nbar = 5$.  From left to right, the three curves correspond to
		drive detunings of $\delta = 3 \gamma$, $\delta = 5 \gamma$ and $\delta = 7 \gamma$.  One sees that increasing the magnitude of the detuning
		causes the negativity to persist to higher temperatures.}	
	\label{fig:IntNeg}
\end{figure}

In Fig.~(\ref{fig:PmTemperature}), we plot the distribution $P(m)$ as obtained from Eq.~(\ref{eq:MGF}) for a large detuning 
$\delta = 10 \gamma$ and for various bath temperatures; as the temperature is lowered, the distribution fails to be positive for 
values of $m > \langle m \rangle$.
Fig.~(\ref{fig:IntNeg}) plots the integrated negativity of the distribution, $N[P] = - \int dm P(m) \theta(-P(m))$ as a function of temperature.  One clearly sees that increasing the drive detuning causes the negativity to emerge at progressively higher temperatures.

In the large detuning limit of interest, one can show analytically (see Appendix B) that it is indeed the anomalously large magnitude of the third cumulant (second term in Eq.~(\ref{eq:QuantumThirdMoment})) which causes the distribution to become non-positive definite.  As shown in Appendix B, 
the large value of the third cumulant causes the distribution to have the form of an Airy function convolved with a Gaussian; the oscillations of the Airy function cause the $P(m)$ to drop below $0$ for values of $m \geq \langle m \rangle$.


\section{Interpretation}
\label{sec:Interpretation}

The interpretation of negative quasi-probabilities of the kind found here (negative ``counting statistics") was first given by Nazarov and Kindermann \cite{Nazarov03b}.  We begin by quickly summarizing their findings, and then extend their interpretation to argue that negative counting statistics correspond to the same sort of non-classical behaviour tested by Leggett-Garg inequalities; in particular, they imply that a ``macrorealistic" classical picture of the fluctuations of $n(t)$ is not possible.

\subsection{Infinite mass detector}
\label{subsec:InfMassDetector}

Nazarov and Kindermann considered an alternate idealized setup for measuring $\hm$, where the detector is an infinitely heavy mass, and the quantity to be measured (in our case $\hn$) is linearly coupled to the position $\hx$ of the detector: 
\begin{eqnarray}
	H_{\rm int} = -A \hat{x} \hat{n}.
	\label{eq:HintInfMass}
\end{eqnarray}
The detector-oscillator interaction is turned on for a time $t$.  Classically, the interaction would simply shift the detector momentum an amount $A n t = A m$, while (due to its infinite mass) its position would be unchanged.  Correspondingly, one would expect that in the quantum case, the final momentum distribution of the detector mass would be a convolution of the detector's initial momentum distribution and the desired probability distribution $P(m)$.
The only additional complication is that there may be a backaction effect:   the distribution $P(m)$ may itself (via $H_{\rm int}$) depend on the value of the detector mass position $x$.  In our case the interaction Hamiltonian in Eq.~(\ref{eq:HintInfMass}) implies that different values of $x$ correspond to different values of the resonator frequency (and hence drive detuning $\delta$).  As a result, different values of $x$ will lead to different distributions $P(m; \delta = \delta_0 + Ax)$.
Including this effect, one would then expect the following relation between the detector mass Wigner function $W(x,p;t)$ before and after the interaction:
\begin{eqnarray}
	W(x,p;t) & = & \int dm P(m; \delta = \delta_0 + A x) W(x,p-A m; 0). \nonumber \\
\end{eqnarray} 
This relation was rigorously derived in Ref.~\cite{Nazarov03b}, with $P(m, \delta)$ being the usual Keldysh-ordered distribution we have been considering.

Consider the case where the detector mass 
is initially in a Gaussian state with zero means, a momentum variance $\sigma_p = A \sigma_{\rm imp}$ and a position variance $\sigma_x = \hbar \sigma_{\rm BA} / A $.  We could then use the final momentum distribution of the mass to infer the distribution $P(m)$ as:
\begin{eqnarray}
	P_{\rm meas}(m; \delta) & \equiv & 
		A \int dx W[x,m/A; t] \nonumber \\
	& = &
		 \frac{1}{ \sqrt{4 \pi^2 \sigma_{\rm imp} \sigma_{\rm BA} } } \int dm'  \int d \delta'
		P[m'; \delta']  \times
	\nonumber  \\
	&&
			\exp \left( -\frac{ (m-m')^2}{2 \sigma_{\rm imp}^2 } \right) \cdot
			\exp \left( -\frac{ (\delta - \delta')^2}{2 \sigma_{\rm BA}^2 } \right). 
			\nonumber \\
			\label{eq:MeasuredDist}
\end{eqnarray}
The above equation provides us with an unambiguous way to interpret $P(m)$. It tells us that the Keldysh ordered $P(m, \delta)$ should be regarded as the ``intrinsic" distribution describing the fluctuations of $m$ for a given fixed value of $\delta$.  In contrast, the measured distribution $P_{\rm meas}(m, \delta)$ is this intrinsic distribution corrupted by the addition of measurement uncertainty.  There is both an imprecision uncertainty $\sigma_{\rm imp}$ in $m$ coming from the momentum uncertainty in the detector mass initial state, and a backaction uncertainty $\sigma_{\rm BA}$ in $\delta$ coming from the detector mass position uncertainty.  The Heisenberg uncertainty principle applied to the detector mass implies that $\sigma_{\rm imp} \sigma_{\rm BA} \geq 1 / 2$: one can never eliminate both these sources of measurement uncertainty.  Nonetheless, Eq.~(\ref{eq:MeasuredDist}) gives us a way to define the underlying, measurement-noise free distribution $P(m)$.

Turning to the issue of positivity, we must have that the measured distribution $P_{\rm meas}(m,\delta)$ is positive definite, as it is just the final momentum distribution of the test mass.  In the special case where $P(m)$ is independent of $x$ (i.e.~a true QND measurement where there is no backaction), this constraint immediately implies that $P(m)$ be positive definite.  However, in the case relevant here, where backaction is important (i.e.~different values of $x$ affect the fluctuations of $m$), this is no longer required:  $P(m)$ can exhibit negativity in these cases without violating the positivity of 
$P_{\rm meas}(m)$.  This is precisely what we we find in the driven number-fluctuation statistics at low temperature and large drive detuning.  

For our driven cavity, $P(m,\delta)$ will only change appreciably when $\delta$ is varied an amount $\sim \gamma$.  To avoid a sizeable backaction, we would thus want $\sigma_{\rm BA} \ll \gamma$.  The Heisenberg uncertainty principle then implies $\sigma_{\rm imp} \gg 1 / \gamma$, which sets a limit to the scale $\Delta m$ of any negative regions in $P(m, \delta)$.  These constraints are indeed obeyed by our results.   

\subsection{Significance of negative probabilities}

With Eq.~(\ref{eq:MeasuredDist}) in hand, we can now view the failure of $P(m)$ to be positive definite as a clear manifestation of non-classical behaviour in our driven resonator.  Classically, we would naturally think of the fluctuations of $m$ in terms of random trajectories $n(t)$ and a corresponding distribution function.  The failure of the Keldysh ordered $P(m)$ to be positive means that even in the most highly idealized measurement setups, we cannot interpret the measured statistics in this way:  they do not correspond to having added measurement noise to an underlying classical stochastic process.  This interpretation can only be salvaged if one is willing to accept that the intrinsic distribution function describing the fluctuations is not positive definite.  

Not surprisingly, negativity in $P(m)$ and the corresponding non-classical behaviour only emerges at sufficiently low temperatures.  More subtle however is the dependence on drive detuning:  negativity only occurs at a sufficiently large-magnitude detuning $|\delta|$, and is always enhanced by increasing $|\delta|$.  On a basic level, this is consistent with  Eq.~(\ref{eq:MeasuredDist}), which tells us that $P(m)$ can only be negative if there is a backaction effect associated with measuring $m$.  In our case, this backaction effect vanishes to leading order when $\delta = 0$ \cite{Clerk10a}.  Hence, it is reasonable that obtaining negativity requires a non-zero detuning $\delta$.

We stress that this non-classicality discussed here is very different than that associated with a non-positive definite Wigner function; here, we are not characterizing the instantaneous state of a system, but rather the time-integrated fluctuations of an observable.  There is rather a much more natural connection to the kind of non-classical temporal correlations that lead to violations of Leggett-Garg inequalities (LGIs) \cite{Leggett85}.  These inequalities constrain temporal correlations of a given observable $O(t)$ in any classical theory which satisfies:
\begin{enumerate}
\item Macrorealism:  $O(t)$ has a definite value at all times.
\item Noninvasive measurability:  $O(t)$ can be measured without any backaction disturbance that would alter its subsequent evolution.
\end{enumerate}
We note that two recent experiments have reported violation of an LGI \cite{Esteve10b, Pryde11}.

In some sense, the non-classicality associated with the negativity of $P(m)$ is stronger than that associated with the violation of an LGI.  Violating an LGI could simply be interpreted as indicating that the measurement is indeed invasive (i.e.~there is backaction), without necessarily indicating a failure of macrorealism.  In contrast, negativity of $P(m)$ tells us more than simply ``backaction exists".  It tells us that a natural way of including backaction effects classically (as additional measurement noise which smears an intrinsic probability distribution, c.f.~Eq.~(\ref{eq:MeasuredDist})) is impossible.  Note that
in our system, backaction effects remain important when measuring $P(m)$ at non-zero detuning even in the more classical regime of high-temperatures; nonetheless, the distribution exhibits no negativity here.  

It also interesting to note that while standard LGIs involve two-time correlation functions, the non-classical behaviour found here only manifests itself when one considers higher-order correlation functions.  Recall that the second moment given in Eq.~(\ref{eq:SecondMoment}) has a completely classical form, where the third moment and higher have non-trivial quantum corrections stemming from the $k$-dependence of the effective thermal occupancy factor $\neff[k]$.  Finally, the standard violation of an LGI involves a qubit undergoing Larmor oscillations \cite{Leggett85}.  Similarly, in our driven resonator negative probabilities only emerge for large-magnitude drive detunings $| \delta | \gtrsim \gamma$, a regime where correlation functions of $\hn(t)$ have a strong oscillatory behaviour.

\section{Measurement issues}
\label{sec:Measurement}

We end with a discussion of how one might experimentally detect evidence of the non-classical photon and phonon number fluctuations described in this paper.  


\subsection{Qubit plus resonator measurement}

One approach would be to experimentally implement the model of Sec.~\ref{subsec:QubitModel}, where a qubit is coupled dispersively to the photon number operator $\hn$ of a driven cavity (c.f.~Eq.(\ref{eq:HInt})). The evolution of the qubit phase (i.e.~its off-diagonal density matrix element $\rho_{\ua \da}$) at various values of the dispersive coupling $k$ directly yields the cumulant generating function of the desired distribution $P(m)$ (c.f.~Eq.~(\ref{eq:TLSKeldyshDefn})).  Such a measurement could be contemplated in a cavity QED or circuit QED systems, where a two-level system (atom or superconducting qubit) is coupled to an electromagnetic cavity.  The required phase information could be extracted by using a Ramsey-interference technique, similar to the seminal experiments of Refs.~\cite{Guerlin07, Gleyzes07}.  Unlike those experiments, the focus here is very different:  the goal is learn about the way the cavity photon number fluctuates over a time $t \gg 1 / \gamma$, as opposed to probe its instantaneous value.

In order to detect evidence of the non-classical photon number fluctuations discussed here, it would be sufficient to see that the third moment $\langle \langle m^3 \rangle \rangle$ is negative.  Given the dispersive qubit-cavity interaction, the odd moments of $P(m)$ will contribute in the long-time limit $t \gg 1 / \gamma$ to the ac-Stark shift $\Delta \Omega_{\rm qb}$ of the qubit frequency.  One has:
\begin{eqnarray}
	\Delta \Omega_{\rm qb} & = & \lim_{t \ra \infty} \left(
		k \frac{\langle m \rangle}{t} - \frac{k^3}{6} \frac{ \langle \langle m^3 \rangle \rangle}{t}  \right) + \mathcal{O}(k^5) \nonumber \\
	& = &
	k \nbar + \frac{k^3}{\gamma^2} \nbar \chi + \mathcal{O}(k^5 \chi^3).
	\label{eq:DeltaOmegaQB}
\end{eqnarray}
In the second line, we have assumed that the driven cavity is in the interesting regime of zero temperature and strong detuning (i.e.~$\chi \ll 1$) where we expect strong quantum effects.   We see that anomalous negative skewness of $P(m)$ manifests itself in the sign of the $k^3$ term in the ac-Stark shift of the qubit frequency.

To resolve this non-linear contribution to the ac-Stark shift, one requires sufficient qubit coherence.  As we must allow the qubit phase to evolve long enough both to be in the long-time limit of the photon-number fluctuations, and to resolve the $k^3$ stark shift, we need that the total dephasing rate
(including the contribution from $T_1$ processes) satisfy:
\begin{eqnarray}
	\Gamma_{\varphi}  < \min \left( \gamma, \frac{k^3}{\gamma^2} \nbar \chi \right).
	\label{eq:DephasingLimit}
\end{eqnarray}
One unavoidable contribution to $\Gamma_{\varphi}$ will come from the dispersive coupling and the even moments of $P(m)$; it is easy to see that this contribution satisfies Eq.~(\ref{eq:DephasingLimit}) as long as one has a strong dispersive coupling $k \gtrsim \gamma$.  This regime has been achieved in several recent circuit QED experiments \cite{Schuster07, Johnson10}.  One also needs the intrinsic, non-cavity dephasing of the qubit to be sufficiently small.  Given recent advances in extending the coherence of superconducting qubits \cite{SchoelkopfPP11}, this also would appear to be feasible.

An alternate approach would be to measure the order $k^3$ term in the ac-Stark shift via simple spectroscopy, where one directly drives the qubit and measures its state as a function of this drive frequency.  In order to avoid complications arising from the spectroscopy drive modifying the cavity state (and hence $P(m)$), one would need to use, e.g., a second cavity for the spectroscopy \cite{Schuster07}.

\subsection{Measurement with a generic linear-response detector}

Perhaps a more general way to measure the fluctuations of $m$ would be to weakly couple the photon (or phonon) number operator $\hn$ we wish to measure to the input port of a generic linear detector \cite{ClerkRMP}, as discussed in Ref.~\cite{Clerk10a}.  We would thus have a coupling of the form:
\begin{eqnarray}
	H_{\rm int, lin} = \hbar \hn \cdot \hF
\end{eqnarray}
where $\hF$ is a detector operator and a generalized force.  We would then monitor some other detector observable, say $\hI$, whose value depends linearly on $\hn$.  The dispersively-coupled optomechanical setup for detecting phonon shot noise analyzed in Ref.~\cite{Clerk10a} falls into this general description.  In that case, $\hF$ is the photon number operator of the optical cavity used to detect mechanical quanta, and $\hI$ is the homodyne current.

In this sort of generic setup, the statistics of the detector output $I$ can be used to to extract the statistics of $m$.  Of course 
(similar to the idealized detector of Sec.~\ref{subsec:InfMassDetector}), this correspondence will be hindered by the presence of both measurement imprecision noise (i.e.~the intrinsic fluctuations in $\hI$) as well as backaction noise (i.e.~the effective fluctuations in detuning resulting from fluctuations in $\hF$).  The simplest evidence for non-classical fluctuations of $m$ comes from the anomalous sign of the third moment; we thus need to ask whether measurement imprecision and backaction would obscure the intrinsic negativity of the skewness.

To that end, we first note that measurement imprecision here can be treated as an additive Gaussian noise process, and hence will not change the value of the third moment.
As for the backaction fluctuations, they are equivalent to having phase fluctuations on the mechanical drive.  Treating these backaction phase fluctuations along the same lines as Ref.~\cite{Rabl09}, we find that they yield an additional additive contribution to $\langle \langle m^3 \rangle \rangle$ which is always positive, and which in the large $\nbar$ limit takes the form:
\begin{eqnarray}
	\langle \langle m^3 \rangle \rangle_{\rm BA} & = &
		\left(  \frac{S_{FF} \nbar}{\gamma} \right)^2 
		\frac{96 \nbar t}{\gamma^2}
		\frac{ \left[ 1 - 12 (\delta / \gamma)^2 \right]^2 }{(1 + 4 \delta^2 / \gamma^2)^4}
\end{eqnarray}
By choosing a sufficiently small measurement strength (i.e.~$S_{FF}$) and large enough detuning, one can still have the total third moment be negative.
In the large detuning limit, the backaction-induced skewness scales as $1/\delta^4$, whereas the intrinsic, negative skewness scales as $1/\delta^2$.
Note also that the backaction contribution to the second moment $\langle \langle m^2 \rangle \rangle$ for this generic linear-detector setup was discussed in Ref.~\cite{Clerk10a}; there, one finds that the minimum possible total added noise is achieved for $S_{FF} \propto 1 / \nbar$.

\section{Conclusions}
\label{sec:Conclusions}

In this paper, we have shown that the full-counting statistics of energy fluctuations in a driven quantum resonator can become negative for sufficiently low temperature and large drive detuning.  This negativity arises from the same kind of quantum correction that leads to a negative third moment \cite{Clerk10a}, something that is impossible classically.  We have argued that the failure of the quasi-probability distribution describing $P(m)$ to be positive definite is similar to having violated a Leggett-Garg inequality, and implies that a ``macrorealistic", classical picture for the fluctuations of phonon/photon number is not possible.
We have thus shown in a relatively simple setting that higher moments of such counting statistics can be used to detect non-classical behaviour.  It would be extremely interesting to investigate whether similar effects manifest themselves in other system.

\section*{Acknowledgements}
We thank T. Bhattacharya for a useful discussion.  This work was supported by the Canadian Institute for Advanced Research and by the DARPA ORCHID program through a grant from AFOSR.


\begin{appendix}

\section{Keldysh ordering from homodyne measurement theory}

Consider the measurement setup of recent optomechanical experiments \cite{Harris08, Sankey10}, where the number operator $\hn(t)$ of a mechanical mode is coupled dispersively (strength $g$) to a driven optical mode which acts as a detector cavity.  The coupling takes the form:
\begin{eqnarray}
	H_{\rm int} = g \hat{n} \ha^{\dag} \ha.
\end{eqnarray}
where $\ha$ is the annihilation operator for the measurement cavity.  By virtue of this interaction, the frequency of the detector cavity will depend on the value of $n(t)$.  One can thus measure the time variation of $n(t)$ by detecting the resulting variation in the detector cavity frequency via homodyne detection of its output field (see, e.g., \cite{WallsMilburn08, ClerkRMP}).  This involves first mixing the detector cavity output field with a large, classical reference beam.  To leading order in $g$,  the output field
$\hb$ from the mixer will have the form:
\begin{eqnarray}
	\hb(t) = \beta + B \hn(t).
	\label{eq:OutputField}
\end{eqnarray}
where $\beta$ parameterizes the large magnitude of the classical reference beam used, $B \propto g$ and we have omitted vacuum noise terms responsible for shot noise (i.e.~the imprecision noise in this measurement scheme).  Without loss of generality, we take both $\beta$ and $B$ to be real.  The intensity $\hI = \hb^\dag \hb$ of the mixer output is then measured using a photodetector.  Assuming that the constant term $\beta^2$ intensity can be subtracted from $\hI$ (by, e.g., using balanced homodyne detection), the output of the detector to leading order in $\beta$ is:
\begin{eqnarray}
	\delta \hI = \hI - \beta^2 \simeq 2 B \beta \hat{n}.
\end{eqnarray}
We see that in the large $\beta$ limit, the output is just linearly proportional to $\hat{n}$.

Given this simple linear correspondence, one can directly infer the value of the $j$th moment of $m$ from the measured $j$th-order intensity correlation function, i.e.
\begin{eqnarray}
	\langle m^j \rangle_{\rm meas} \equiv
	\lim_{\beta \ra \infty}
		\frac{1}{(2 \beta B)^j} \int_0^t \prod_{l=1}^j d t'_l \langle \tilde{\mathcal{T}} 
			\left[  \delta \hI(t'_1) ... \hI(t'_j) \right] \rangle. \nonumber \\
		\label{eq:NOTO}
\end{eqnarray} 
On the RHS of this expression, the symbol $\tilde{\mathcal{T}}$ denotes that the measured intensity correlation functions correspond to expectation values which are both normal-ordered and time-ordered with respect to the $b(t)$ and $b^{\dag}(t)$ operators; this ordering prescription is a direct consequence of measuring intensity via photodetection \cite{Glauber63, Carmichael87}.  For example, for the third moment we have:
\begin{eqnarray}
	\tilde{\mathcal{T}} \langle I(t_1) I(t_2) I(t_3) \rangle  = 
		\langle \hb^\dag(t_a) \hb^\dag(t_b) \hb^\dag(t_c) \hb(t_c) \hb(t_b) \hb(t_a) \rangle,
		\nonumber \\
\end{eqnarray}		
where $t_a < t_b < t_c$ denotes the time-ordered listing of $t_1,t_2$ and $t_3$.  

Eq.~(\ref{eq:NOTO}) and Eq.~(\ref{eq:OutputField}) completely determine the correspondence between the measured moments $\langle m^j \rangle_{\rm meas}$ and appropriately ordered expectation values of products of $\hm(t)$.  Though tedious, one can now explicitly confirm that for each moment $\langle m^j \rangle_{\rm meas}$, the resulting ordering of $\hm(t)$ operators is exactly the Keldysh ordering defined by Eq.~(\ref{eq:TLSKeldyshDefn}).  For the third moment, this was done in Ref.~\cite{Clerk10a}.

A more compact way to see that one obtains the Keldysh ordering at each order is to use a functional-integral formulation of the Keldysh technique; a pedagogical introduction to this approach is given in Refs.~\cite{Kamenev05, Kamenev09}.  
In this formulation, each bosonic operators is replaced by two time-dependent fields, 
e.g.~$\hb(t) \ra b_\sigma(t), \hn(t) \ra n_\sigma(t) $, where the index $\sigma = +(-)$ denotes the forward (backwards) Keldysh contour.  Within this approach,
different operator orderings correspond to different combinations of $+$ and $-$ fields.  A special role is played by the so-called ``classical" field, which is the 
average of $+$ and $-$ fields, i.e. 
\begin{eqnarray}
	n_{\rm cl}(t) \equiv \frac{n_{+}(t) + n_{-}(t)}{2}
\end{eqnarray}
At the level of a saddle-point approximation, the dynamics of the classical field correspond to an effective classical Langevin equation.
Correlation functions of this classical field are obtained in the usual way using the Keldysh action $\mathcal{S}$ describing the system:
\begin{eqnarray}
	\langle n_{\rm cl}(t_1)... n_{\rm cl}(t_j) \rangle \equiv
		\int \prod_{j,\sigma = \pm} \mathcal{D} \phi_{j\sigma} n_{\rm cl}(t_1)... n_{\rm cl}(t_j) 
		\exp[ i \mathcal{S}] .
		\nonumber \\
\end{eqnarray}
The $\phi_{j \sigma}(t)$ here denote the various fields that describe the system;  the action is a function of these fields.  By construction, the $j$th-order correlation functions defined above is identical to the $j$th order, Keldysh-ordered operator expectation value defined by Eq.~(\ref{eq:TLSKeldyshDefn})
\cite{Kamenev05}.

Turning to our homodyne measurement, we first note that the normal-ordered, time-ordered intensity correlation functions that are measured via photodetection can be obtained by adding an auxiliary source term to the Keldysh action $\mathcal{S}$ of the form:
\begin{eqnarray}
	\mathcal{S}_{src} = k \int_0^t dt' \left[  b_{-}^*(t) b_{+} (t) - \beta^2 \right].
\end{eqnarray}
Given the correspondence between the Keldysh $\pm$ fields and operator orderings \cite{Kamenev09}, 
one finds that derivatives of the full Keldysh partition function 
(action $\mathcal{S} + \mathcal{S}_{src}$) with respect to $k$ at $k=0$ generate the desired normal and time-ordered correlation functions in the usual way.  

Next, for homodyne detection, we can make the replacement:
\begin{eqnarray}
	b_\sigma(t) & \simeq & b^*_\sigma(t) \simeq \beta + B n_\sigma(t), 
\end{eqnarray}
which results in:
\begin{eqnarray}
	\mathcal{S}_{src} & \simeq & 
		2 k \beta B \int_0^t dt'   n_{\rm cl}(t),
\end{eqnarray}
where we have only retained the leading-order-in-$\beta$ term in the action.  We thus see that the source field $k$ couples to the classical field $n_{\rm cl}(t)$; it thus follows that the $j$th-order intensity correlation functions (as generated by differentiation of the Keldysh partition function with respect to $k$) will be directly proportional to $j$th order Keldysh-ordered correlation functions of $m$.


\section{$P(m)$ in the large time, large detuning limit}
\label{appendix:Airy}

We first shift and rescale $P(m)$ in the full quantum case so that it has zero mean and unit variance.  
Setting $\nth=0$, the CGF $\tilde{\Lambda}_{\rm dr}[k]$ of the transformed distribution takes the form:
\begin{eqnarray}
	\tilde{\Lambda}_{\rm dr}[k] & = & 
		-\frac{k^2}{2} \left(
			\frac{
				1 
				+ i \frac{1}{4  \sqrt{\chi \nbar t  \gamma} } k
				}{1+
					i \sqrt{ \frac{\chi}{\nbar t  \gamma} }k - \frac{1}{4 \nbar t  \gamma} k^2}
		\right).
\end{eqnarray}

Consider now the strong-detuning, long-time limit, such that $\nbar \gamma t \ra \infty$ but $\chi \nbar \gamma t $ is finite.  In this limit
\begin{eqnarray}
	\tilde{\Lambda}_{\rm dr}[k] & \ra &
			-\frac{1}{2}k^2 
			-i \frac{1}{  8 \sqrt{\chi \nbar t \gamma } } k^3.
	\label{eq:ScaledCGF}
\end{eqnarray}
In the quantum case, both the second and third moments are non-vanishing in this long-time, strong-detuning limit.  In contrast, in the same limit the classical distribution would be completely Gaussian.  Thus, the third moment term in Eq.~(\ref{eq:ScaledCGF}) is entirely due to the effective $k$ dependence of the thermal factor $\neff$ in the quantum distribution.  Further, note that increasing the drive detuning (and hence reducing $\chi$) {\it enhances} the non-Gaussian nature of the distribution described by Eq.~(\ref{eq:ScaledCGF}); this is the opposite of what happens classically, where a large detuning suppresses fluctuations and non-Gaussian effects, as the magnitude of thermal fluctuations at the drive frequency are reduced.


Fourier transforming the approximate CGF in Eq.~(\ref{eq:ScaledCGF}) reveals that in the large detuning limit, the distribution $\Pdr(m)$ is a convolution of a Gaussian and an Airy function.  This can be explicitly evaluated.  Defining $\tilde{m} = (m - \nbar t) / \sqrt{\langle \langle m^2 \rangle \rangle}$), we have:
\begin{eqnarray}
	P(\tilde{m},t) & \simeq & \frac{1}{ \lambda }  \exp \left[
		\frac{-1}{2 \lambda^3} \left( \tm -
			\frac{1}{6 \lambda^3} \right) \right]
			\textrm{Ai }\left[ -\frac{\tm}{\lambda} + \frac{1}{4  \lambda^4} \right],
			\nonumber \\ 
			\label{eq:AiryDist}
\end{eqnarray}
where 
\begin{eqnarray}
	\lambda^3 & = & -\frac{  \langle \langle m^3 \rangle \rangle }{2  \langle \langle m^2 \rangle \rangle ^{3/2} } \sim
			 \frac{3}{  8 \sqrt{\chi \nbar t \gamma } } .
\end{eqnarray}
It is the oscillation of the Airy function factor in Eq.~(\ref{eq:AiryDist}) above which gives rise to the negative probabilities at $\tm > 0$.  The exponential prefactor ensures that the resulting negativity is exponentially suppressed in the long time limit when $\lambda \ll 1$.  However, for intermediate times (still much longer than $1/\gamma$), one has $\lambda \gtrsim 1$, and the negativity can be appreciable.  Note that the most prominent domain of negativity in this large-$\lambda$ limit has an extent in $\tm \sim \lambda$; in terms of $m/t$, this corresponds to a range $< 1$.  Thus, while $\Pdr(m)$ exhibits negativity even in a seemingly classical regime where $\nbar \gamma t \gg 1$, it only occurs on a scale which corresponds to less than one quantum in the resonator.  This is consistent with the discussion of negative quasi-probabilities given in Sec.~\ref{subsec:InfMassDetector}.


\end{appendix}

\bibliographystyle{apsrev}
\bibliography{ACTotalRefsV2}

\begin{thebibliography}{30}
\expandafter\ifx\csname natexlab\endcsname\relax\def\natexlab#1{#1}\fi
\expandafter\ifx\csname bibnamefont\endcsname\relax
  \def\bibnamefont#1{#1}\fi
\expandafter\ifx\csname bibfnamefont\endcsname\relax
  \def\bibfnamefont#1{#1}\fi
\expandafter\ifx\csname citenamefont\endcsname\relax
  \def\citenamefont#1{#1}\fi
\expandafter\ifx\csname url\endcsname\relax
  \def\url#1{\texttt{#1}}\fi
\expandafter\ifx\csname urlprefix\endcsname\relax\def\urlprefix{URL }\fi
\providecommand{\bibinfo}[2]{#2}
\providecommand{\eprint}[2][]{\url{#2}}

\bibitem[{\citenamefont{Grangier et~al.}(1998)\citenamefont{Grangier, Levenson,
  and Poizat}}]{Grangier98}
\bibinfo{author}{\bibfnamefont{P.}~\bibnamefont{Grangier}},
  \bibinfo{author}{\bibfnamefont{J.~A.} \bibnamefont{Levenson}},
  \bibnamefont{and} \bibinfo{author}{\bibfnamefont{J.-P.}
  \bibnamefont{Poizat}}, \bibinfo{journal}{Nature}
  \textbf{\bibinfo{volume}{396}}, \bibinfo{pages}{537} (\bibinfo{year}{1998}).

\bibitem[{\citenamefont{Nogues et~al.}(1999)\citenamefont{Nogues,
  Rauschenbeutel, Osnaghi, Brune, Raimond, and Haroche}}]{Haroche99}
\bibinfo{author}{\bibfnamefont{G.}~\bibnamefont{Nogues}},
  \bibinfo{author}{\bibfnamefont{A.}~\bibnamefont{Rauschenbeutel}},
  \bibinfo{author}{\bibfnamefont{S.}~\bibnamefont{Osnaghi}},
  \bibinfo{author}{\bibfnamefont{M.}~\bibnamefont{Brune}},
  \bibinfo{author}{\bibfnamefont{J.~M.} \bibnamefont{Raimond}},
  \bibnamefont{and} \bibinfo{author}{\bibfnamefont{S.}~\bibnamefont{Haroche}},
  \bibinfo{journal}{Nature (London)} \textbf{\bibinfo{volume}{400}},
  \bibinfo{pages}{239} (\bibinfo{year}{1999}).

\bibitem[{\citenamefont{Guerlin et~al.}(2007)\citenamefont{Guerlin, Bernu,
  Deleglise, Sayrin, Gleyzes, Kuhr, Brune, Raimond, and Haroche}}]{Guerlin07}
\bibinfo{author}{\bibfnamefont{C.}~\bibnamefont{Guerlin}},
  \bibinfo{author}{\bibfnamefont{J.}~\bibnamefont{Bernu}},
  \bibinfo{author}{\bibfnamefont{S.}~\bibnamefont{Deleglise}},
  \bibinfo{author}{\bibfnamefont{C.}~\bibnamefont{Sayrin}},
  \bibinfo{author}{\bibfnamefont{S.}~\bibnamefont{Gleyzes}},
  \bibinfo{author}{\bibfnamefont{S.}~\bibnamefont{Kuhr}},
  \bibinfo{author}{\bibfnamefont{M.}~\bibnamefont{Brune}},
  \bibinfo{author}{\bibfnamefont{J.-M.} \bibnamefont{Raimond}},
  \bibnamefont{and} \bibinfo{author}{\bibfnamefont{S.}~\bibnamefont{Haroche}},
  \bibinfo{journal}{Nature} \textbf{\bibinfo{volume}{448}},
  \bibinfo{pages}{889} (\bibinfo{year}{2007}).

\bibitem[{\citenamefont{Gleyzes et~al.}(2007)\citenamefont{Gleyzes, Kuhr,
  Guerlin, Bernu, Deleglise, Hoff, Brune, Raimond, and Haroche}}]{Gleyzes07}
\bibinfo{author}{\bibfnamefont{S.}~\bibnamefont{Gleyzes}},
  \bibinfo{author}{\bibfnamefont{S.}~\bibnamefont{Kuhr}},
  \bibinfo{author}{\bibfnamefont{C.}~\bibnamefont{Guerlin}},
  \bibinfo{author}{\bibfnamefont{J.}~\bibnamefont{Bernu}},
  \bibinfo{author}{\bibfnamefont{S.}~\bibnamefont{Deleglise}},
  \bibinfo{author}{\bibfnamefont{U.~B.} \bibnamefont{Hoff}},
  \bibinfo{author}{\bibfnamefont{M.}~\bibnamefont{Brune}},
  \bibinfo{author}{\bibfnamefont{J.-M.} \bibnamefont{Raimond}},
  \bibnamefont{and} \bibinfo{author}{\bibfnamefont{S.}~\bibnamefont{Haroche}},
  \bibinfo{journal}{Nature} \textbf{\bibinfo{volume}{446}},
  \bibinfo{pages}{297} (\bibinfo{year}{2007}).

\bibitem[{\citenamefont{Schuster et~al.}(2007)\citenamefont{Schuster, Houck,
  Schreier, Wallraff, Gambetta, Blais, Frunzio, Johnson, Devoret, Girvin
  et~al.}}]{Schuster07}
\bibinfo{author}{\bibfnamefont{D.}~\bibnamefont{Schuster}},
  \bibinfo{author}{\bibfnamefont{A.}~\bibnamefont{Houck}},
  \bibinfo{author}{\bibfnamefont{J.}~\bibnamefont{Schreier}},
  \bibinfo{author}{\bibfnamefont{A.}~\bibnamefont{Wallraff}},
  \bibinfo{author}{\bibfnamefont{J.}~\bibnamefont{Gambetta}},
  \bibinfo{author}{\bibfnamefont{A.}~\bibnamefont{Blais}},
  \bibinfo{author}{\bibfnamefont{L.}~\bibnamefont{Frunzio}},
  \bibinfo{author}{\bibfnamefont{B.}~\bibnamefont{Johnson}},
  \bibinfo{author}{\bibfnamefont{M.}~\bibnamefont{Devoret}},
  \bibinfo{author}{\bibfnamefont{S.}~\bibnamefont{Girvin}},
  \bibnamefont{et~al.}, \bibinfo{journal}{Nature}
  \textbf{\bibinfo{volume}{445}}, \bibinfo{pages}{515} (\bibinfo{year}{2007}).

\bibitem[{\citenamefont{Johnson et~al.}(2010)\citenamefont{Johnson, Reed,
  Houck, Schuster, Bishop, Ginossar, Gambetta, Dicarlo, Frunzio, Girvin
  et~al.}}]{Johnson10}
\bibinfo{author}{\bibfnamefont{B.~R.} \bibnamefont{Johnson}},
  \bibinfo{author}{\bibfnamefont{M.~D.} \bibnamefont{Reed}},
  \bibinfo{author}{\bibfnamefont{A.~A.} \bibnamefont{Houck}},
  \bibinfo{author}{\bibfnamefont{D.~I.} \bibnamefont{Schuster}},
  \bibinfo{author}{\bibfnamefont{L.~S.} \bibnamefont{Bishop}},
  \bibinfo{author}{\bibfnamefont{E.}~\bibnamefont{Ginossar}},
  \bibinfo{author}{\bibfnamefont{J.~M.} \bibnamefont{Gambetta}},
  \bibinfo{author}{\bibfnamefont{L.}~\bibnamefont{Dicarlo}},
  \bibinfo{author}{\bibfnamefont{L.}~\bibnamefont{Frunzio}},
  \bibinfo{author}{\bibfnamefont{S.}~\bibnamefont{Girvin}},
  \bibnamefont{et~al.}, \bibinfo{journal}{Nature Phys.}
  \textbf{\bibinfo{volume}{6}}, \bibinfo{pages}{1} (\bibinfo{year}{2010}).

\bibitem[{\citenamefont{Thompson et~al.}(2008)\citenamefont{Thompson, Zwickl,
  Jayich, Marquardt, Girvin, and Harris}}]{Harris08}
\bibinfo{author}{\bibfnamefont{J.~D.} \bibnamefont{Thompson}},
  \bibinfo{author}{\bibfnamefont{B.~M.} \bibnamefont{Zwickl}},
  \bibinfo{author}{\bibfnamefont{A.~M.} \bibnamefont{Jayich}},
  \bibinfo{author}{\bibfnamefont{F.}~\bibnamefont{Marquardt}},
  \bibinfo{author}{\bibfnamefont{S.~M.} \bibnamefont{Girvin}},
  \bibnamefont{and} \bibinfo{author}{\bibfnamefont{J.~G.~E.}
  \bibnamefont{Harris}}, \bibinfo{journal}{Nature (London)}
  \textbf{\bibinfo{volume}{452}}, \bibinfo{pages}{72} (\bibinfo{year}{2008}).

\bibitem[{\citenamefont{Sankey et~al.}(2010)\citenamefont{Sankey, Yang, Zwickl,
  Jayich, and Harris}}]{Sankey10}
\bibinfo{author}{\bibfnamefont{J.~C.} \bibnamefont{Sankey}},
  \bibinfo{author}{\bibfnamefont{C.}~\bibnamefont{Yang}},
  \bibinfo{author}{\bibfnamefont{B.~M.} \bibnamefont{Zwickl}},
  \bibinfo{author}{\bibfnamefont{A.~M.} \bibnamefont{Jayich}},
  \bibnamefont{and} \bibinfo{author}{\bibfnamefont{J.~G.~E.}
  \bibnamefont{Harris}}, \bibinfo{journal}{Nature Phys.}
  (\bibinfo{year}{2010}).

\bibitem[{\citenamefont{Clerk et~al.}(2010{\natexlab{a}})\citenamefont{Clerk,
  Marquardt, and Harris}}]{Clerk10a}
\bibinfo{author}{\bibfnamefont{A.~A.} \bibnamefont{Clerk}},
  \bibinfo{author}{\bibfnamefont{F.}~\bibnamefont{Marquardt}},
  \bibnamefont{and} \bibinfo{author}{\bibfnamefont{J.~G.~E.}
  \bibnamefont{Harris}}, \bibinfo{journal}{Phys. Rev. Lett.}
  \textbf{\bibinfo{volume}{104}}, \bibinfo{pages}{213603}
  (\bibinfo{year}{2010}{\natexlab{a}}).

\bibitem[{\citenamefont{Levitov and Lesovik}(1993)}]{Levitov93}
\bibinfo{author}{\bibfnamefont{L.}~\bibnamefont{Levitov}} \bibnamefont{and}
  \bibinfo{author}{\bibfnamefont{G.}~\bibnamefont{Lesovik}},
  \bibinfo{journal}{Jetp Letters} \textbf{\bibinfo{volume}{58}},
  \bibinfo{pages}{230} (\bibinfo{year}{1993}).

\bibitem[{\citenamefont{Levitov et~al.}(1996)\citenamefont{Levitov, Lee, and
  Lesovik}}]{Levitov96}
\bibinfo{author}{\bibfnamefont{L.}~\bibnamefont{Levitov}},
  \bibinfo{author}{\bibfnamefont{H.}~\bibnamefont{Lee}}, \bibnamefont{and}
  \bibinfo{author}{\bibfnamefont{G.}~\bibnamefont{Lesovik}},
  \bibinfo{journal}{J. Math. Phys.} \textbf{\bibinfo{volume}{37}},
  \bibinfo{pages}{4845} (\bibinfo{year}{1996}).

\bibitem[{\citenamefont{Nazarov and Kindermann}(2003)}]{Nazarov03b}
\bibinfo{author}{\bibfnamefont{Y.~V.} \bibnamefont{Nazarov}} \bibnamefont{and}
  \bibinfo{author}{\bibfnamefont{M.}~\bibnamefont{Kindermann}},
  \bibinfo{journal}{Eur. Phys. J. B} \textbf{\bibinfo{volume}{35}},
  \bibinfo{pages}{413} (\bibinfo{year}{2003}).

\bibitem[{\citenamefont{Nagaev}(2002)}]{Nagaev02}
\bibinfo{author}{\bibfnamefont{K.}~\bibnamefont{Nagaev}},
  \bibinfo{journal}{Phys. Rev. B} \textbf{\bibinfo{volume}{66}}
  (\bibinfo{year}{2002}).

\bibitem[{\citenamefont{Beenakker et~al.}(2003)\citenamefont{Beenakker,
  Kindermann, and Nazarov}}]{Beenakker03}
\bibinfo{author}{\bibfnamefont{C.~W.~J.} \bibnamefont{Beenakker}},
  \bibinfo{author}{\bibfnamefont{M.}~\bibnamefont{Kindermann}},
  \bibnamefont{and} \bibinfo{author}{\bibfnamefont{Y.~V.}
  \bibnamefont{Nazarov}}, \bibinfo{journal}{Phys. Rev. Lett.}
  \textbf{\bibinfo{volume}{90}} (\bibinfo{year}{2003}).

\bibitem[{\citenamefont{Belzig and Nazarov}(2001)}]{Belzig01}
\bibinfo{author}{\bibfnamefont{W.}~\bibnamefont{Belzig}} \bibnamefont{and}
  \bibinfo{author}{\bibfnamefont{Y.~V.} \bibnamefont{Nazarov}},
  \bibinfo{journal}{Phys. Rev. Lett.} \textbf{\bibinfo{volume}{87}},
  \bibinfo{pages}{197006} (\bibinfo{year}{2001}).

\bibitem[{\citenamefont{Leggett and Garg}(1985)}]{Leggett85}
\bibinfo{author}{\bibfnamefont{A.}~\bibnamefont{Leggett}} \bibnamefont{and}
  \bibinfo{author}{\bibfnamefont{A.}~\bibnamefont{Garg}},
  \bibinfo{journal}{Phys. Rev. Lett} \textbf{\bibinfo{volume}{54}},
  \bibinfo{pages}{857} (\bibinfo{year}{1985}).

\bibitem[{\citenamefont{Pilgram and B{\"u}ttiker}(2003)}]{Pilgram03}
\bibinfo{author}{\bibfnamefont{S.}~\bibnamefont{Pilgram}} \bibnamefont{and}
  \bibinfo{author}{\bibfnamefont{M.}~\bibnamefont{B{\"u}ttiker}},
  \bibinfo{journal}{Phys. Rev. B} \textbf{\bibinfo{volume}{67}}
  (\bibinfo{year}{2003}).

\bibitem[{\citenamefont{Gardiner and Zoller}(2000)}]{Gardiner00}
\bibinfo{author}{\bibfnamefont{C.~W.} \bibnamefont{Gardiner}} \bibnamefont{and}
  \bibinfo{author}{\bibfnamefont{P.}~\bibnamefont{Zoller}},
  \emph{\bibinfo{title}{Quantum Noise}} (\bibinfo{publisher}{Springer, Berlin},
  \bibinfo{year}{2000}).

\bibitem[{\citenamefont{Clerk et~al.}(2010{\natexlab{b}})\citenamefont{Clerk,
  Devoret, Girvin, Marquardt, and Schoelkopf}}]{ClerkRMP}
\bibinfo{author}{\bibfnamefont{A.~A.} \bibnamefont{Clerk}},
  \bibinfo{author}{\bibfnamefont{M.~H.} \bibnamefont{Devoret}},
  \bibinfo{author}{\bibfnamefont{S.~M.} \bibnamefont{Girvin}},
  \bibinfo{author}{\bibfnamefont{F.}~\bibnamefont{Marquardt}},
  \bibnamefont{and} \bibinfo{author}{\bibfnamefont{R.~J.}
  \bibnamefont{Schoelkopf}}, \bibinfo{journal}{Rev. Mod. Phys.}
  \textbf{\bibinfo{volume}{82}}, \bibinfo{pages}{1155}
  (\bibinfo{year}{2010}{\natexlab{b}}).

\bibitem[{\citenamefont{Glauber}(1963)}]{Glauber63}
\bibinfo{author}{\bibfnamefont{R.~J.} \bibnamefont{Glauber}},
  \bibinfo{journal}{Phys. Rev.} \textbf{\bibinfo{volume}{130}},
  \bibinfo{pages}{2529} (\bibinfo{year}{1963}).

\bibitem[{\citenamefont{Kamenev}(2005)}]{Kamenev05}
\bibinfo{author}{\bibfnamefont{A.}~\bibnamefont{Kamenev}}, in
  \emph{\bibinfo{booktitle}{Nanophysics: Coherence and Transport}}, edited by
  \bibinfo{editor}{\bibfnamefont{H.}~\bibnamefont{Bouchiat}}
  \bibnamefont{et~al.} (\bibinfo{publisher}{Elsevier, Amsterdam},
  \bibinfo{year}{2005}), p. \bibinfo{pages}{177}.

\bibitem[{\citenamefont{Kamenev and Levchenko}(2009)}]{Kamenev09}
\bibinfo{author}{\bibfnamefont{A.}~\bibnamefont{Kamenev}} \bibnamefont{and}
  \bibinfo{author}{\bibfnamefont{A.}~\bibnamefont{Levchenko}},
  \bibinfo{journal}{Advances in Physics} \textbf{\bibinfo{volume}{58}},
  \bibinfo{pages}{197} (\bibinfo{year}{2009}).

\bibitem[{\citenamefont{Clerk and Utami}(2007)}]{Clerk07}
\bibinfo{author}{\bibfnamefont{A.~A.} \bibnamefont{Clerk}} \bibnamefont{and}
  \bibinfo{author}{\bibfnamefont{D.}~\bibnamefont{Utami}},
  \bibinfo{journal}{Phys. Rev. A} \textbf{\bibinfo{volume}{75}},
  \bibinfo{pages}{042302} (\bibinfo{year}{2007}).

\bibitem[{\citenamefont{Widder}(1946)}]{HamburgerRef}
\bibinfo{author}{\bibfnamefont{D.~V.} \bibnamefont{Widder}},
  \emph{\bibinfo{title}{The Laplace Transform}} (\bibinfo{publisher}{Princeton
  University Press, Princeton}, \bibinfo{year}{1946}).

\bibitem[{\citenamefont{Palacios-Laloy
  et~al.}(2010)\citenamefont{Palacios-Laloy, Mallet, Nguyen, Bertet, Vion,
  Esteve, and Korotkov}}]{Esteve10b}
\bibinfo{author}{\bibfnamefont{A.}~\bibnamefont{Palacios-Laloy}},
  \bibinfo{author}{\bibfnamefont{F.}~\bibnamefont{Mallet}},
  \bibinfo{author}{\bibfnamefont{F.}~\bibnamefont{Nguyen}},
  \bibinfo{author}{\bibfnamefont{P.}~\bibnamefont{Bertet}},
  \bibinfo{author}{\bibfnamefont{D.}~\bibnamefont{Vion}},
  \bibinfo{author}{\bibfnamefont{D.}~\bibnamefont{Esteve}}, \bibnamefont{and}
  \bibinfo{author}{\bibfnamefont{A.~N.} \bibnamefont{Korotkov}},
  \bibinfo{journal}{Nat. Phys.} \textbf{\bibinfo{volume}{6}},
  \bibinfo{pages}{442} (\bibinfo{year}{2010}).

\bibitem[{\citenamefont{Goggin et~al.}(2011)\citenamefont{Goggin, Almeida,
  Barbieri, Lanyon, O'Brien, White, and Pryde}}]{Pryde11}
\bibinfo{author}{\bibfnamefont{M.~E.} \bibnamefont{Goggin}},
  \bibinfo{author}{\bibfnamefont{M.~P.} \bibnamefont{Almeida}},
  \bibinfo{author}{\bibfnamefont{M.}~\bibnamefont{Barbieri}},
  \bibinfo{author}{\bibfnamefont{B.~P.} \bibnamefont{Lanyon}},
  \bibinfo{author}{\bibfnamefont{J.~L.} \bibnamefont{O'Brien}},
  \bibinfo{author}{\bibfnamefont{A.~G.} \bibnamefont{White}}, \bibnamefont{and}
  \bibinfo{author}{\bibfnamefont{G.~J.} \bibnamefont{Pryde}},
  \bibinfo{journal}{Proc. Acad. Nat. Sci.} \textbf{\bibinfo{volume}{108}},
  \bibinfo{pages}{1256} (\bibinfo{year}{2011}).

\bibitem[{\citenamefont{Paik et~al.}(2011)}]{SchoelkopfPP11}
\bibinfo{author}{\bibfnamefont{H.}~\bibnamefont{Paik}} \bibnamefont{et~al.},
  \bibinfo{journal}{arXiv:1105.4652v1}  (\bibinfo{year}{2011}).

\bibitem[{\citenamefont{Rabl and Aspelmeyer}(2009)}]{Rabl09}
\bibinfo{author}{\bibfnamefont{P.}~\bibnamefont{Rabl}} \bibnamefont{and}
  \bibinfo{author}{\bibfnamefont{M.}~\bibnamefont{Aspelmeyer}},
  \bibinfo{journal}{Physical Review A} \textbf{\bibinfo{volume}{80}}
  (\bibinfo{year}{2009}).

\bibitem[{\citenamefont{Walls and Milburn}(2008)}]{WallsMilburn08}
\bibinfo{author}{\bibfnamefont{D.~F.} \bibnamefont{Walls}} \bibnamefont{and}
  \bibinfo{author}{\bibfnamefont{G.~J.} \bibnamefont{Milburn}},
  \emph{\bibinfo{title}{Quantum Optics}} (\bibinfo{publisher}{Springer,
  Berlin}, \bibinfo{year}{2008}), \bibinfo{edition}{2nd} ed.

\bibitem[{\citenamefont{Carmichael}(1987)}]{Carmichael87}
\bibinfo{author}{\bibfnamefont{H.}~\bibnamefont{Carmichael}},
  \bibinfo{journal}{J. Opt. Soc. Am. B} \textbf{\bibinfo{volume}{4}},
  \bibinfo{pages}{1588} (\bibinfo{year}{1987}).

\end{thebibliography}
\end{document}